\documentclass[aps,amsmath,amssymb,twocolumn,prl,superscriptaddress]{revtex4-2}

\usepackage{graphicx}
\usepackage{dcolumn}
\usepackage{bm}
\usepackage{color}
\usepackage[normalem]{ulem}
\usepackage{amsmath}
\usepackage{hyperref}

\begin{document}

\title{Quartet Tomography in Multiterminal Josephson Junctions}

\author{David Christian Ohnmacht}
\email{david.ohnmacht@uni-konstanz.de}
\affiliation{Fachbereich Physik, Universit{\"a}t Konstanz, D-78457 Konstanz, Germany}

\author{Marco Coraiola}
\affiliation{IBM Research Europe---Zurich, Säumerstrasse 4, 8803 Rüschlikon, Switzerland}

\author{Juan José García-Esteban}
\affiliation{Fachbereich Physik, Universit{\"a}t Konstanz, D-78457 Konstanz, Germany}
\affiliation{Departamento de F\'{i}sica Te\'{o}rica de la Materia Condensada and Condensed Matter Physics Center (IFIMAC), 
Universidad Aut\'{o}noma de Madrid, E-28049 Madrid, Spain}

\author{Deividas Sabonis}
\affiliation{IBM Research Europe---Zurich, Säumerstrasse 4, 8803 Rüschlikon, Switzerland}

\author{Fabrizio Nichele}
\affiliation{IBM Research Europe---Zurich, Säumerstrasse 4, 8803 Rüschlikon, Switzerland}

\author{Wolfgang Belzig}
\affiliation{Fachbereich Physik, Universit{\"a}t Konstanz, D-78457 Konstanz, Germany}

\author{Juan Carlos Cuevas}
\email{juancarlos.cuevas@uam.es}
\affiliation{Fachbereich Physik, Universit{\"a}t Konstanz, D-78457 Konstanz, Germany}
\affiliation{Departamento de F\'{i}sica Te\'{o}rica de la Materia Condensada and Condensed Matter Physics Center (IFIMAC), 
Universidad Aut\'{o}noma de Madrid, E-28049 Madrid, Spain}

\date{\today}

\begin{abstract}
We investigate the detection of quartets in hybrid multiterminal Josephson junctions. Using simple models of quantum 
dots coupled to superconducting leads, we find that quartets are ubiquitous in quantum coherent structures 
and show how to rigorously extract their contribution to the current-phase relation (CPR). We also demonstrate 
that quartets are closely related to the hybridization of Andreev bound states (ABSs) in these systems and propose a method 
to identify quartets directly in ABS spectra. We illustrate our method by analyzing the spectroscopic measurements of the 
ABS spectrum in a three-terminal Josephson junction realized in an InAs/Al heterostructure. Our analysis strongly suggests 
the existence of quartets in the studied hybrid system.
\end{abstract}

\maketitle

\emph{Introduction}.-- A junction with a short normal conducting region between two superconductors accommodates ABSs 
which carry a supercurrent through the system \cite{Beenakker1991,Furusaki1991}. 
A unique aspect of ABSs lies in the tunability of their energies by changing the superconducting phase difference, which has 
led to the proposal of ABSs as a platform for quantum computing \cite{Desposito2001,Zazunov2003,Chtchelkatchev2003,Padurariu2010}. 
Additionally, there have been numerous studies reporting on the CPR and the energy spectra of ABSs in two-terminal Josephson 
junctions (JJs) \cite{Pillet2010,Chang2013,Bretheau2013,Bretheau2013a,Janvier2015,Bretheau2017,vanWoerkom2017,Hays2018,Tosi2019,
Nichele2020,Hays2020,Hays2021,Pita-Vidal2023,Hinderling2023,Wesdorp2022}. In multiterminal JJs (MTJJs), the ABSs energies depend 
on multiple phase differences, enabling band structure engineering which is of particular interest for realizing topological 
systems \cite{Yokoyama2015,Riwar2016,Eriksson2017,Meyer2017,Xie2017,Xie2019,Repin2019,PeraltaGavensky2019,Houzet2019,Klees2020,
Weisbrich2021,Xie2022,Barakov2023,Teshler2023}. 

MTJJs exhibit a plethora of unique transport phenomena. For instance, in a three-terminal device the commensuration 
of voltages at two terminals induces dc supercurrents at finite bias, known as voltage-induced Shapiro steps \cite{Cuevas2007}. 
The same signature was predicted to be related to a transport mechanism referred to as quartet process, which involves one 
Andreev reflection at terminal $i$ and $j$, respectively, and two crossed Andreev reflections at terminal $k$, resulting in 
a dependence of the supercurrent on a combination of phases given by $\varphi_i + \varphi_j - 2\varphi_k$ (where $\varphi_\alpha$ 
indicates the superconducting phase of terminal $\alpha$, for $\alpha = i,j,k$) \cite{Freyn2011,Jonckheere2013}. Additional 
theoretical work has proposed strategies to identify quartets both in the current-voltage characteristics 
\cite{Melin2016,Melin2017,Melin2019,Jacquet2020,Doucot2020,Melin2020,Melin2020a,Melin2021,Melin2022,Melin2023,Jonckheere2023} 
and in the supercurrent \cite{Rech2014,Feinberg2015,Melo2022,Melin2023a}. Following experimental studies resulted in the observation
of potential signatures of these processes obtained by probing the system response as a function of two bias currents 
or voltages \cite{Pfeffer2014,Cohen2018,Huang2022,Arnault2022}. Other experiments in hybrid MTJJs, not specifically designed to 
detect quartets, have also reported signatures compatible with their existence \cite{Draelos2019,Graziano2020,Pankratova2020,
Arnault2021,Matsuo2022,Zhang2023}. A related system, based on two JJs in proximity of each other, has been proposed to realize 
Andreev molecules, where hybridization of ABSs occurs \cite{Pillet2019,Kornich2019,Kornich2020,Pillet2020}. In this 
case, signs of a unconventional coupling mechanism have been predicted and they have been investigated via 
current-bias measurements \cite{Matsuo2022,Haxell2023,Matsuo2023a}. In addition, spectroscopic evidence of ABS hybridization 
was recently reported in a hybrid three-terminal JJ, where ABS spectra were probed as a function of two phase differences 
\cite{Coraiola2023}. However, despite all experimental efforts and theoretical predictions, the unambiguous detection of 
quartet processes is still a challenge.

In this Letter, we present a scheme which we term \textit{Quartet Tomography} to rigorously extract the quartet contributions to the CPR or the ABS spectrum of a quantum coherent MTJJ. We apply this scheme to recent measurements
of the ABS spectrum of a three-terminal JJ realized in an InAs/Al heterostructure \cite{Coraiola2023} and demonstrate the existence 
of quartet processes in these MTJJs. Additionally, we show that the quartet contributions are intimately related to the degree 
of hybridization of the ABSs in these heterostructures and that ABS hybridization is a sufficient, but not necessary condition 
for the existence of quartet processes.


%
\begin{figure*}[t]
\includegraphics[width=0.95\textwidth,clip]{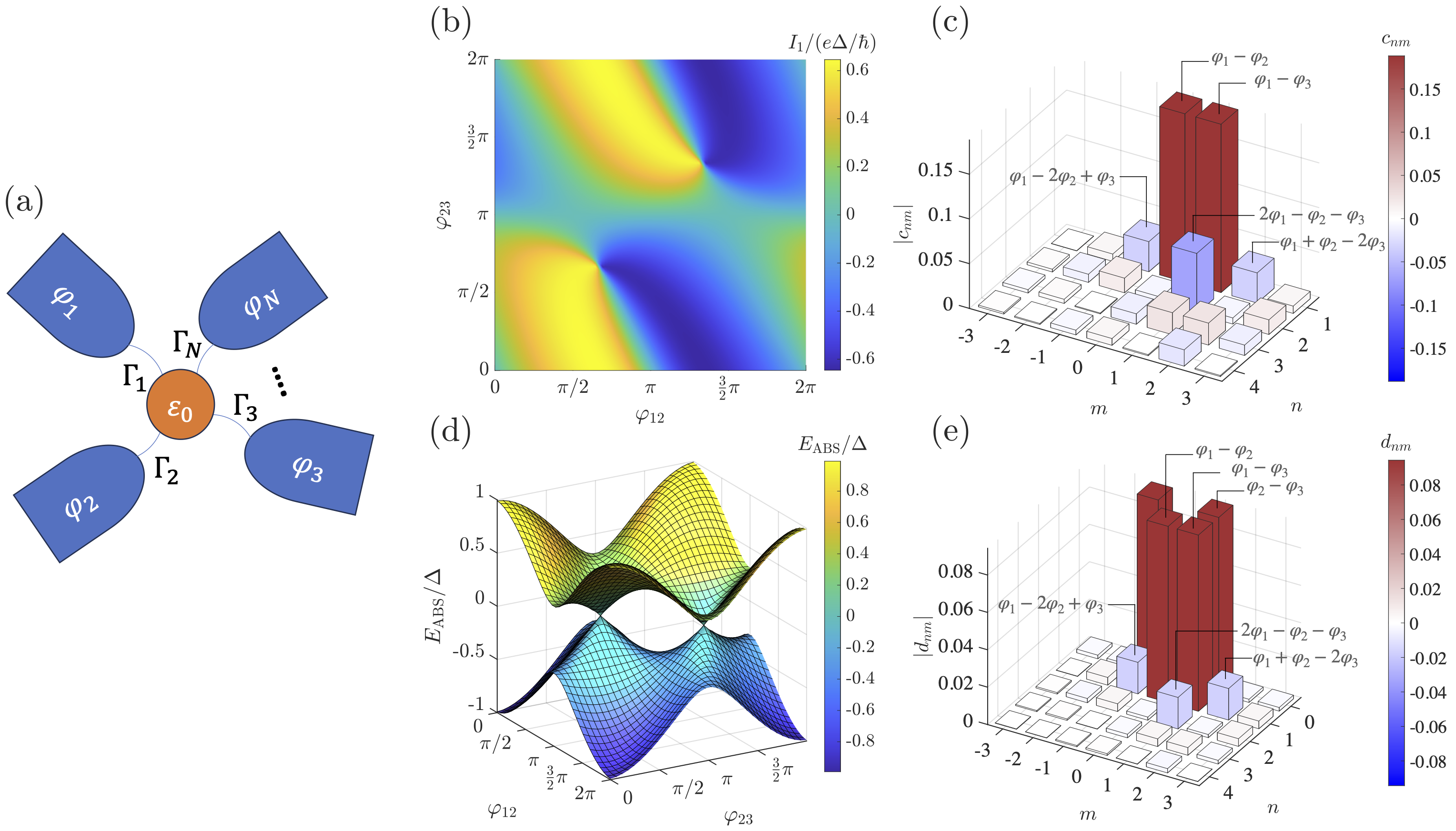}
\caption{(a) Schematics of the single-dot model. A single-level quantum dot is coupled to $N$ superconducting terminals with 
phases $\varphi_j$. The parameter $\Gamma_j$ describes the strength of the coupling between the dot and lead $j$. (b) Example 
of the current-phase relation $I_1(\varphi_{12},\varphi_{23})$ for the model in (a) with three identical leads with energy 
gap $\Delta$. The different parameters are: $\Gamma_{1} = \Gamma_{2} = \Gamma_{3} = 5\Delta$ and $\epsilon_0 = 0$. 
(c) The corresponding $c_{nm}$ coefficients in the expansion of Eq.~\eqref{eq_I1_1D} in units of $e\Delta / \hbar$.
The phase dependence associated to some of these coefficients is indicated in the graph. (d) Andreev bound state spectrum 
$E_{\rm ABS}(\varphi_{12},\varphi_{23})$ for the example shown in panel (b). (e) The corresponding $d_{nm}$ coefficients in the 
expansion of Eq.~\eqref{eq-ABS-phase} in units of $\Delta$.}
\label{fig-1Dot-ex}
\end{figure*}

\emph{Quartet Tomography: single-dot model}.-- To illustrate the idea of quartets and how they show up in the CPR of a 
MTJJ we consider the model illustrated in Fig.~\ref{fig-1Dot-ex}(a), where a single noninteracting quantum dot is coupled 
to $N$ superconducting leads. The quantum dot has a single spin-degenerate level of energy $\epsilon_0$ and the superconducting
electrodes are $s$-wave superconductors characterized by energy gaps $\Delta_j$ and superconducting phases $\varphi_j$
($j=1,\dots,N$). The coupling between the dot and lead $j$ is described by the tunneling rate $\Gamma_j$. Our goal is to compute 
the supercurrent flowing in the terminals of this system as a function of the phases $\varphi_j$. For this purpose, we 
employ Green's function techniques and our input are the (dimensionless) retarded and advanced Green's functions of the leads, 
which in a $2 \times 2$ Nambu representation read $\hat{g}^{\rm r/a}_j(E) = g^{\rm r/a}_j \hat{\tau}_0 + f^{\rm r/a}_j 
e^{\imath \varphi_j \hat{\tau}_3}\hat{\tau}_1$, with $g^{\rm r/a}_j = -(E\pm \imath \eta)/\sqrt{\Delta_j^2-(E\pm \imath \eta)^2}$ 
and $f^{\rm r/a}_j =-\Delta_j/\sqrt{\Delta_j^2-(E\pm \imath \eta)^2}$. Here, $E$ is the energy, $\eta = 0^+$, and $\tau_{0,1}$
are Pauli matrices. As we show in the Supplemental Material \cite{SM}, one can express the current flowing 
through lead $i$ as $I_i = \sum_{j \neq i} I_{ij}$, where $I_{ij}$ is given by
\begin{equation} \label{eq_Iij}
    I_{ij} = \frac{8e}{h} \Gamma_i \Gamma_j \sin(\varphi_{ji}) \int^{\infty}_{-\infty} \Im \left\{ 
    \frac{f^{\rm a}_i(E) f^{\rm a}_j(E)}{D(E,\vec{\varphi})} \right\} n_{\rm F}(E) \, dE .
\end{equation}
Here, $\varphi_{ij} = \varphi_i - \varphi_j$, $n_{\rm F}(E)$ is the Fermi function, $\vec{\varphi} = 
(\varphi_1, \dots, \varphi_N)$, and $D(E,\vec{\varphi})$ is given by
\begin{eqnarray} \label{eq-D}
D(E,\vec{\varphi}) & = & \left[ E - \epsilon_0 - \sum_k \Gamma_k g^{\rm a}_k \right]  
\left[ E + \epsilon_0 - \sum_k \Gamma_k g^{\rm a}_k \right] \nonumber \\ 
& & - \left[ \sum_k \Gamma_k f^{\rm a}_k e^{\imath \varphi_k} \right]  
\left[ \sum_k \Gamma_k f^{\rm a}_k e^{-\imath \varphi_k} \right] .
\end{eqnarray}
The ABS energies in this structure are derived from the condition $D(E,\vec{\varphi}) = 0$.

We focus now on a three-terminal device and consider the current $I_1$. 
This current depends on two phase differences, chosen $\varphi_{12} = \varphi_1 - \varphi_2$ 
and $\varphi_{23} = \varphi_2 - \varphi_3$, and can be expressed as the Fourier series:
\begin{equation} \label{eq_I1_1D}
    I_1(\varphi_{12},\varphi_{23}) = \sum_{n,m} c_{nm} \sin (n \varphi_{12} + m \varphi_{23}) ,
\end{equation}
with $c_{0m} = 0$, namely there are no contributions of the type $\sin(\varphi_2 - \varphi_3)$, 
and $c_{nm} = -c_{-n-m}$.

In this system, a quartet is a correlated Cooper pair tunneling process that involves three terminals and whose 
contribution to the supercurrent depends on a phase of the type $\varphi^{\rm q}_k = \varphi_i + \varphi_j - 2 \varphi_k$ 
with $i,j \neq k$ and $i \neq j$. There are three types of quartets, and from Eq.~\eqref{eq_Iij} one can show that 
the corresponding contributions to leading order in the $\Gamma$ parameters are given by \cite{SM}: $c_{-2,-1} = 2Q_1, 
c_{1,-1} = -Q_{2}, c_{1,2} = -Q_3$, where
\begin{eqnarray} \label{eq_Qk_1D}
    Q_k & = & \frac{8e}{h} \Gamma_i \Gamma_j \Gamma^2_k \int^{\infty}_{-\infty}dE \, n_{\rm F}(E) \times \\
    & & \Im \left\{ \frac{f^{\rm a}_i(E) f^{\rm a}_j(E) [f^{\rm a}_k(E)]^2}{(E^2 - \epsilon^2_0)^2} 
    \right\} . \nonumber
\end{eqnarray}
Equation~\eqref{eq_Qk_1D} supports the interpretation that a quartet of the type
$\varphi^{\rm q}_1 = \varphi_2 + \varphi_3 - 2 \varphi_1$ involves the injection of two Cooper pairs from terminal 1 that
are transferred separately to leads 2 and 3. Let us emphasize that the supercurrent $I_1(\varphi_{12},\varphi_{23})$
contains not only quartet contributions of the kind $\sin (\varphi^{\rm q}_k)$, but also harmonics of this phase. More
importantly, Eq.~\eqref{eq_I1_1D} suggests a direct way to extract the quartet contributions, which consists in performing 
a Fourier analysis of the CPR. This idea was proposed in Ref.~\cite{Rech2014} in the context of a
three-terminal system including four junctions with a carbon nanotube. Here, we show that it holds in quantum coherent
structures irrespective of the details of the normal scattering region. We illustrate the results for this model 
in Fig.~\ref{fig-1Dot-ex} where we show both the CPR $I_{\rm 1}(\varphi_{12},\varphi_{23})$ in panel (b) and the $c_{nm}$ 
coefficients in the expansion of Eq.~\eqref{eq_I1_1D} in panel (c). We note the appearance of the three types of quartet 
contributions predicted above, see panel (c). Moreover, as expected from Eq.~\eqref{eq_Qk_1D} for a symmetric situation (all
$\Gamma$'s and $\Delta$'s equal), the quartet $\varphi^{\rm q}_1$ has a magnitude that is twice that of the other two, 
$\varphi^{\rm q}_{2}$ and $\varphi^{\rm q}_{3}$, while it contributes to the current $I_1$ with an opposite sign. It is 
worth stressing that quartet contributions are accompanied by other, more dominant contributions related to terms proportional 
to $\sin(\varphi_{ij})$ and their harmonics. Those additional contributions originate from the tunneling of single and multiple 
Cooper pairs, respectively (see Ref.~\cite{SM}). Their unavoidable presence complicates the identification of quartets in the 
analysis of transport properties such as the critical current. This problem is resolved by our quartet tomography because it 
does not rely on the relative magnitude of the different contributions.


\emph{ABS energies and supercurrent}.-- As we have just shown, quartets are easily identified from the CPR. 
However, to isolate the CPR of one terminal in a phase-controlled MTJJ is challenging. For this reason, we propose 
a second method based on the measurement of the density of states (DOS) by means of tunneling spectroscopy like in 
Ref.~\cite{Coraiola2023}. From the DOS we can deduce the ABS spectrum, which in turn is closely related to the supercurrent.
In our MTJJ, the zero-temperature current $I_1$ is obtained from the energies $E^{(l)}_{\rm ABS}$ of the occupied ABSs as
follows
\begin{equation} \label{eq-ABS-super}
	I_1(\vec{\varphi}) = \frac{2e}{\hbar} \sum_l \frac{\partial E^{(l)}_{\rm ABS}(\vec{\varphi})}{\partial \varphi_1} .
\end{equation}  
The factor 2 appears because we are assuming spin degeneracy. We have verified that, in the examples shown
in this work, we can reconstruct the CPR using Eq.~\eqref{eq-ABS-super}. This demonstrates that the supercurrent 
is carried by the ABSs with no contributions from the continuum. Therefore, there is an even simpler protocol to identify 
quartets, which consists in the Fourier analysis of the ABS spectrum. In the case of a three-terminal junction, this 
spectrum admits a Fourier expansion of the type:
\begin{equation} \label{eq-ABS-phase}
    E^{(l)}_{\rm ABS}(\vec{\varphi}) = -\sum_{nm} d^{(l)}_{nm} \cos(n\varphi_{12} + m\varphi_{23}).
\end{equation}
The Fourier coefficients $d^{(l)}_{nm}$ are related to the Fourier coefficients of the supercurrent via 
Eq.~\eqref{eq-ABS-super}. The results for the energy of the occupied and unoccupied ABS for the single-dot model are 
shown in Fig.~\ref{fig-1Dot-ex}(d). These two states touch at zero energy in this example because the condition
$D(E,\vec{\varphi}) = 0$, see Eq.~(\ref{eq-D}), is fulfilled for zero energy when the couplings have similar values. 
The corresponding Fourier components for the occupied ABS are displayed in panel (e). Here, the color indicates whether 
a contribution is positive or negative, whereas the bar height is a measure of its absolute value. Notice that quartet 
contributions are clearly visible and in accordance with the Fourier components of the CPR in Fig.~\ref{fig-1Dot-ex}(c). 
Note also that the coefficient $d_{00}$ is not shown because it does not contribute to the supercurrent.

\begin{figure*}[t]
\includegraphics[width=0.95\textwidth,clip]{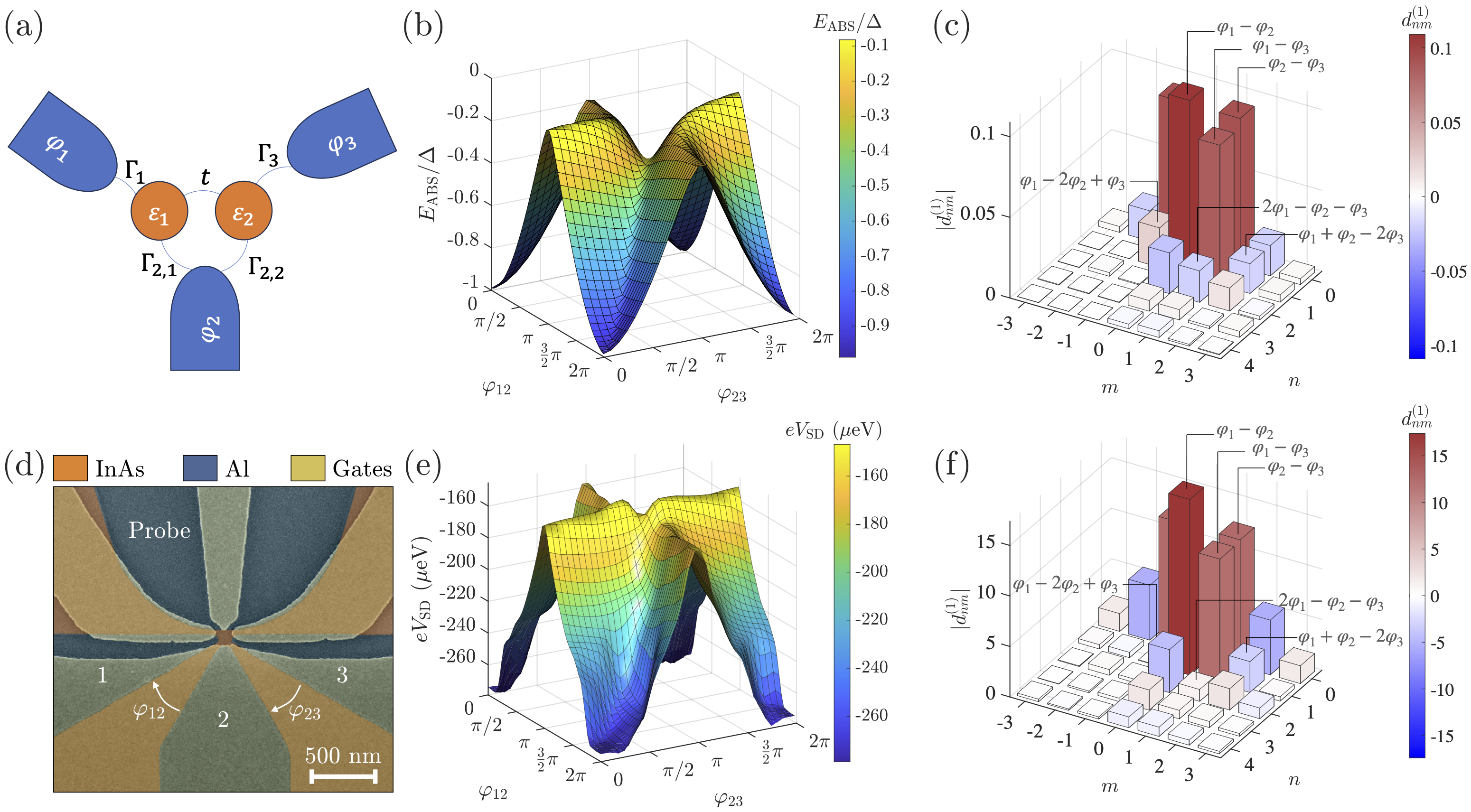}
\caption{(a) Schematics of the double-dot model. (b) Example of the energy $E_{\rm ABS}^{(1)}(\varphi_{12}, 
\varphi_{23})$ of the highest occupied ABS in the model of panel (a). The different parameters are: 
$\Delta_{1,2,3} = \Delta$, $\Gamma_{1} = 5.5\Delta$, $\Gamma_{2,1} = 6\Delta$, $\Gamma_{2,2} = 5\Delta$, 
$\Gamma_{3} = 6\Delta$, $t = 5\Delta$, and $\epsilon_1 = \epsilon_2 = 0$. (c) The corresponding $d^{(1)}_{nm}$ 
coefficients in the expansion of Eq.~\eqref{eq-ABS-phase} measured in units of $\Delta$. The phase dependence 
associated to some of these coefficients is indicated in the graph. (d) False-colored scanning electron micrograph 
of the device measured in Ref.~\cite{Coraiola2023}, near to the three-terminal Josephson junction region. Exposed 
semiconductor is shown in orange, aluminum in blue and gate electrodes in gold. (e) The upper ABS extracted from 
the tunneling spectroscopy data of Ref.~\cite{Coraiola2023}, where $V_{\rm SD}$ is the bias voltage applied to 
the superconducting probe. (f) The corresponding $d^{(1)}_{nm}$ coefficients of Eq.~\eqref{eq-ABS-phase} in $\mu$eV 
that show the presence of quartets in the studied device.}
\label{fig-2Dot-ex}
\end{figure*}

\emph{Quartets and ABS hybridization: double-dot model}.-- Since quartets are correlated tunneling events involving 
three terminals, they are closely related to the hybridization of ABSs in these structures. To illustrate this 
idea we now consider the double quantum dot model illustrated in Fig.~\ref{fig-2Dot-ex}(a). This model 
was recently used to describe the hybridization of ABSs in Ref.~\cite{Coraiola2023} (see below). In this case, the 
normal scattering region is formed by two single-level quantum dots (with energies $\epsilon_1$ and $\epsilon_2$). 
These dots are coupled to three superconducting terminals as shown in Fig.~\ref{fig-2Dot-ex}(a). There is an interdot 
coupling described by the parameter $t$, which controls the degree of hybridization of the ABSs. Again, using Green's 
function techniques one can compute the supercurrent that flows through the different terminals; the details are presented 
in Ref.~\cite{SM}. Crucially, the current $I_1$ admits the same Fourier expansion of Eq.~\eqref{eq_I1_1D}. A 
perturbative analysis shows that one has the same three types of quartets as in the single-dot model, and all 
contributions have a similar form. For instance, for the quartet $\varphi^{\rm q}_{3} = \varphi_{1} + 
\varphi_{2} - 2\varphi_{3}$, we obtain a contribution to the current of the form $Q_{3} \sin(\varphi^{\rm q}_{3})$, 
where $Q_{3}$ is given to leading order in $t$ and the $\Gamma$ parameters [see Fig.~\ref{fig-2Dot-ex}(a)] by
\begin{eqnarray} \label{eq_Qk_2D}
    Q_{3} & = & \frac{8e}{h} t^2 \Gamma_{1} \Gamma^2_{3}  \int^{\infty}_{-\infty} dE \,
    n_{\rm F}(E) \times \\ 
    & & \Im \left\{ \frac{f^{\rm a}_{1}(E) f^{\rm a}_{2}(E) [f^{\rm a}_{3}(E)]^2}
    {(E^2 - \epsilon^2_1) (E^2 - \epsilon^2_2)} \left( \frac{\Gamma_{2,1}}{E^2 - \epsilon^2_1} + 
    \frac{\Gamma_{2,2}}{E^2 - \epsilon^2_2} \right) \right\} . \nonumber
\end{eqnarray}
This expression shows that quartets would not be possible without ABS hybridization, as $Q_3$ vanishes for
zero interdot coupling ($t=0$).

We illustrate the results for this model in Fig.~\ref{fig-2Dot-ex}(b,c) where we show an example of the energy of the 
highest occupied ABS, $E_{\rm ABS}^{(1)}(\varphi_{12},\varphi_{23})$, along with the $d^{(1)}_{nm}$ coefficients in the 
expansion of Eq.~\eqref{eq-ABS-phase} (see caption for parameter values). Hybridization signs of the ABSs connecting the 
pairs of terminals 1-2 and 2-3 are clearly visible in the region $(\varphi_{12},\varphi_{23}) \sim (\pi,\pi)$, where the 
ABS energy differs from that of the two independent states. More importantly, this hybridization gives rise to three quartet
contributions, which again are accompanied by terms that involve phase differences between only two terminals.
We note that because of the presence of two ABSs, there is no one-to-one correspondence between the coefficients of 
the upper ABS $d^{(1)}_{nm}$ and the supercurrent coefficients $c_{nm}$ as there can be cancellations arising from
contributions of the lower ABS $d^{(2)}_{nm}$. However, a nonzero quartet contribution of an ABS always results in 
a nonvanishing quartet contribution to the supercurrent. Thus, it is enough to obverse nonzero quartet contributions 
of the upper ABS to demonstrate their existence. A detailed discussion of this issue and the relation between quartets 
and ABS hybridization is provided in Ref.~\cite{SM}. 


\emph{Quartet Tomography in a three-terminal Josephson junction}.-- The two protocols described above to identify
quartet contributions can be directly applied to experimental data. We now illustrate this with the analysis of the 
quartet contributions in the ABS spectrum measured in Ref.~\cite{Coraiola2023}. The device investigated in that work, 
shown in Fig.~\ref{fig-2Dot-ex}(d), consisted of three superconducting terminals (Al) coupled to a normal region (InAs). 
Terminals 1 and 2 (2 and 3) were connected to form a closed loop, hence enabling control over the phase difference 
$\varphi_{12}$ ($\varphi_{23}$) via integrated flux-bias lines. The DOS in the normal region was probed via tunneling 
spectroscopy, performed by applying a voltage bias $V_{\rm SD}$ to a fourth superconducting terminal. Further 
details about materials, fabrication and measurements can be found in Ref.~\cite{Coraiola2023}. Using the information 
of the tunneling spectra as a function of the two phase differences and with the help of a deep learning algorithm 
\cite{SM}, we reconstructed the phase dependence of the highest occupied ABS in this device, see Fig.~\ref{fig-2Dot-ex}(e). 
Notice that the ABS spectrum is presented in terms of the probe voltage and thus, it is offset by the gap of the
superconducting probe \cite{Coraiola2023}, which has no influence in the tomography. From this ABS spectrum, and using 
the protocol described above, we obtained the coefficients of the quartet tomography $d^{(1)}_{nm}$ that are shown in 
Fig.~\ref{fig-2Dot-ex}(f). Again, the coefficient $d^{(1)}_{00}$ is not shown. Notice that the quartet contributions 
$d^{(1)}_{12}$ and $d^{(1)}_{21}$ related to the phases $\varphi^{\rm q}_3$ and $\varphi^{\rm q}_1$, respectively, 
have a nonzero magnitude resulting in a nonzero quartet contribution to the supercurrent. As mentioned above, 
the specific values of $d^{(1)}_{12/21}$ do not reflect in general the quartet amplitudes in the supercurrent. However,
in our experiments, in which we infer a relatively large hybridization (apparent in the large gap between the two upper
ABSs), the extracted quartet contributions in Fig.~\ref{fig-2Dot-ex}(f) are in fact good estimates for the 
corresponding supercurrent amplitudes, as shown in Ref.~\cite{SM}. More importantly, the quartet contributions do not 
vanish, thus demonstrating their existence in this system. These quartet contributions are a  
consequence of the significant ABS hybridization in this device, which is clearly visible in the region 
$(\varphi_{12},\varphi_{23}) \sim (\pi,\pi)$ and whose signatures were amply discussed in Ref.~\cite{Coraiola2023}. 
This issue is further addressed in Ref.~\cite{SM}, where we also present an analysis of the robustness of 
the quartet tomography.


\emph{Conclusions}.-- We have put forward two protocols to identify quartets in multiterminal Josephson junctions. 
In particular, we have shown how quartet contributions can be extracted from measurements of the ABS spectrum 
of a three-terminal hybrid Josephson junction \cite{Coraiola2023}. We have also shown that in heterostructures
featuring several ABSs, as it is normally the case, quartets are intimately related to the hybridization of ABSs. 
It is worth stressing that the protocol based on the analysis of the CPR could, in principle, be used to identify 
quartets in CPR measurements in Andreev molecules \cite{Haxell2023,Matsuo2023a}. From a theoretical point of view, 
it would be interesting, among other things, to study the connection between quartets and topological states, as well
as to investigate the impact of the spin-orbit interaction in these correlated tunneling events 
\cite{vanHeck2014,Coraiola2023a}.


We thank Aleksandr E.\ Svetogorov for helpful discussions. D.C.O.\ and W.B.\ acknowledge support by the Deutsche 
Forschungsgemeinschaft (DFG; German Research Foundation) via SFB 1432 (Project No. 425217212). F.N.\ acknowledges 
support from the European Research Council (grant number 804273) and the Swiss National Science Foundation (grant 
number 200021\_201082). J.J.G.E.\ and J.C.C.\ were supported by the Spanish Ministry of Science and Innovation through 
a FPU grant (FPU19/05281) and the project No.\ PID2020-114880GB-I00, respectively, and thank the DFG and SFB 1432 
for sponsoring their stay at the University of Konstanz (J.C.C.\ as a Mercator Fellow).

\bibliographystyle{apsrev4-2}
\bibliography{MyLibrary}

\end{document}